\documentstyle[pra,aps,psfig]{revtex}
\begin{document}
\preprint{}
\draft
%
%
\title{Measurement quantum mechanics\\   
and experiments on quantum Zeno effect}
\author{Carlo Presilla}
\address{Dipartimento di Fisica, Universit\`a di Roma ``La Sapienza,''
and INFN, Sezione di Roma,\\
Piazzale A. Moro 2, Roma, Italy 00185}
\author{Roberto Onofrio}
\address{Dipartimento di Fisica ``G. Galilei,'' Universit\`a di Padova,
and INFN, Sezione di Padova,\\
Via Marzolo 8, Padova, Italy 35131}
\author{Ubaldo Tambini}
\address{Dipartimento di Fisica, Universit\`a di Ferrara,
and INFN, Sezione di Ferrara,\\
Via Paradiso 12, Ferrara, Italy 44100}
\date{cond-mat/9603182, Annals of Physics, June 1996}
\maketitle
%
%
\begin{abstract}
Measurement quantum mechanics, the theory of a quantum system which 
undergoes a measurement process, is introduced by a loop of mathematical
equivalencies connecting previously proposed approaches.
The unique phenomenological parameter of the theory is linked to the physical 
properties of an informational environment acting as a measurement apparatus
which allows for an objective role of the observer.
Comparison with a recently reported experiment suggests how
to investigate novel interesting regimes for the quantum Zeno effect. 
\end{abstract}
%
%
%
%
\section{Introduction}

The description of the measurement process has been a topic debated 
from the early developments of quantum mechanics \cite{VONNEUM,WHEE,BUSCH}.
Besides being a basic issue in the interpretation of the quantum 
formalism it has also a practical interest in predicting the results 
of experiments pointing out some paradoxical aspects of the quantum laws 
when compared to the classical ones.
The development of devices whose noise figures are close to the quantum 
limit makes these experiments within the technological feasibility and
demands for a systematization of the theory of quantum
measurement with deeper understanding of its experimental implications
\cite{GREEN,BRAGIN}.

In ordinary quantum mechanics measurements are taken into account 
by postulating the wave function collapse \cite{VONNEUM}.  
This approach has the unpleasant feature of introducing 
an extra assumption in the theory which, moreover, regards only instantaneous 
and perfect measurements. 
In the last twenty years relevant steps have been made in upgrading
the von Neumann postulate with more realistic and satisfactory approaches.
These approaches essentially recognize that a measured system is 
not isolated but in interaction with a measurement apparatus. 
The way this basic fact is taken into account in 
the evolution law of the measured system has been developed according to
different languages and points of view. 

A group algebra approach to the problem of an open quantum system was
proposed by postulating (completely positive) semigroup properties for 
the dynamical law of a system in interaction with a Markovian environment 
\cite{GORINI,LINDBLAD}.
In this case the open system is described by a semigroup master equation 
later successfully used in modeling quantum optics experiments 
\cite{CARMICHAEL}.
The semigroup master equation,
still preserving positivity and trace of the density matrix operator,
introduces decoherentization by dynamically quenching the off-diagonal 
elements of the density matrix, a key property used to explain the absence
of superposition states in a measurement apparatus \cite{ZUREK}
or in a general macroscopic system \cite{GHIRIWE,DIOSI2} 
thought as systems interacting with an environment.
Open quantum systems were gathered to measured systems 
by obtaining an all alike semigroup master equation for 
a continuous measurement process \cite{BALAMPRO} modeled by 
repeated instantaneous effect-valued measurements \cite{LUDWIG}, 
i.e., partial localization or decoherentization kicks \cite{DGHP} given
to the density matrix at random values of the measured observable.

In the model \cite{BALAMPRO} the semigroup master equation comes out 
only after averaging the instantaneous results of the measured 
observable with probability distribution in agreement with standard 
quantum mechanics \cite{BARCHIELLI}.
In absence of this average, i.e., for a particular {\sl selection} of the  
measurement outcomes, a nonlinear stochastic differential 
equation was proposed to describe the evolution of the density matrix during
a continuous measurement process \cite{DIOSI1}. 
Since in this case during the measurement process the density matrix 
coincides with its square, a nonlinear 
stochastic differential equation could be derived for the corresponding 
wave function \cite{DIOSI1B} crowning previous attempts to include 
the L\"uders postulate \cite{LUDER}, a generalization of the von Neumann 
one to {\sl selective} measurements, into a stochastic Schr\"odinger 
equation \cite{GISIN1}.
This equation, obtained also in the framework of the quantum filtering 
theory \cite{BELAVKIN1}, is a special case of the more general
quantum state diffusion equation for open systems \cite{GISIN2}.
Its nonlinearity can be avoided only renouncing to make predictions
on the outcome of the measured observable.
Indeed, a linear stochastic differential equation was 
proposed for the so called {\sl a posteriori} states 
\cite{BELAVKIN2,BARBE}, i.e., those unnormalized states which describe 
the quantum system when the measurement result is already known.

As originally suggested by Feynman \cite{FEY} the path integral approach 
to quantum mechanics is a quite appropriate framework for incorporating
the effect of a measurement.
The hint was picked up by modeling a continuous measurement 
of position having a certain result (selective measurement) by a
restriction of the Feynman path integral \cite{ME1}.
The method was extended to measurements of a generic observable, function 
of momentum and position, by a restriction of the quantum 
propagator in the phase-space formulation \cite{GOMES,OPT}.
The restriction was originally proposed 
as a Gaussian functional damping the paths proportionally to the 
time-averaged squared difference between the value taken by the observable 
on the paths and the measurement result, normalized to a given 
variance \cite{MENSKYBOOK}.
However, the corresponding quantum propagator has desirable dynamical
semigroup properties if the Gaussian restriction is chosen linear in time 
\cite{NAMIOT} which is equivalent to make the previously proposed 
variance scaling with the inverse of the measurement time \cite{TPO}.

The modifications introduced in the quantum propagator by the restricted 
path integral approach can be incorporated into an effective 
Hamiltonian depending on the selected measurement result.  
An effective wave equation was then derived which is 
linear in the wave function if the selected measurement result 
is considered known \cite{MEOP1,MEOP2,CALO}.
On the other hand, if the selected measurement result 
is to be determined according to the evolution of 
the measured system the effective wave equation is read as a 
nonlinear stochastic differential equation. 
Equivalence to the previously proposed stochastic equations occurs
\cite{DIOSI4}.

The selective constraint imposed to the restricted path integral 
approach can be removed by summing over all possible measurement 
outcomes. 
In this way a nonselective process described by a density matrix 
was obtained \cite{MENSKY94}.
The effect of the interaction with the measurement apparatus 
modifies the corresponding quantum propagator through an influence 
functional \cite{FV} which turns out to be equivalent to those  
obtained with other methods \cite{BALAMPRO,CAMIL,CAVES86}.

From the above incomplete list of approaches, 
apart from an evident difference in the languages, 
a rather unified picture of the problem of quantum measurement emerges.
It deserves uprising to the systematic theory of a quantum system evolving 
under the effect of a measurement process.
Such theory, named, for brevity, measurement quantum mechanics,
is formally presented in section II.
 
A less formal introduction to measurement quantum mechanics is attainable 
through the analysis of a model of measurement device which is
rather general.
This is important not only as a justification of the formal approach to
the theory but, above all, for clearly defining the meaning of a measurement
in relation to the observer.
As noted by Cini \cite{CINI} paradoxical features arise from not considering 
the objective role of the observer in a measurement process.
The evolution of a measured system in interaction with a measurement 
apparatus can not depend on the observer looking or not at the
pointer of the apparatus.
In order to take into account such objectivity the measurement
apparatus must be classical with respect to the observer \cite{CINI}.
In section III we propose a simple model of measurement apparatus
shaped as an environment of particles linearly interacting with
the measured system and in contact with a heat reservoir at a fixed
temperature \cite{CALDLEGG}.
In the high temperature limit the particles behave as an 
informational environment which extracts information in objective way.
Noticeably, the formal structure of measurement quantum mechanics 
is obtained in the same limit. 

A further analysis of measurement quantum mechanics
is given in section IV in connection to a recent optical experiment 
\cite{ITA} showing quantum Zeno effect. 
Comparison of the experimental results with the theoretical predictions
indicate that the experiment \cite{ITA} was performed 
in a regime of very strong
coupling of the measurement apparatus with the measured system.
Repetition of this experiment or similar ones in a weaker coupling regime 
is desirable for investigating interesting quantum features.

Some final remarks are given in section V.

\section{Measurement quantum mechanics}

In this section we show the mathematical equivalence of the five approaches 
to the problem of measurement in quantum systems briefly described in the
introduction.
The demonstration is organized in steps relating neighboring pairs of 
approaches and giving rise to the equivalence loop sketched in Fig. 1.
No one of these equivalence steps is novel.  
However, we reconsider all them in a unified framework and language
with naturally emerging definitions for concepts as selective and 
nonselective measurements as well as {\sl a priori} and 
{\sl a posteriori} analysis of a measurement process.
As a result we get a theory
which takes into account the effect of a measurement process in
ordinary quantum dynamics through a phenomenological parameter 
coupling the measured system to the measurement apparatus.
We enter the loop of Fig. 1 at the group algebra approach 
to the master equation and go on in the clockwise sense.
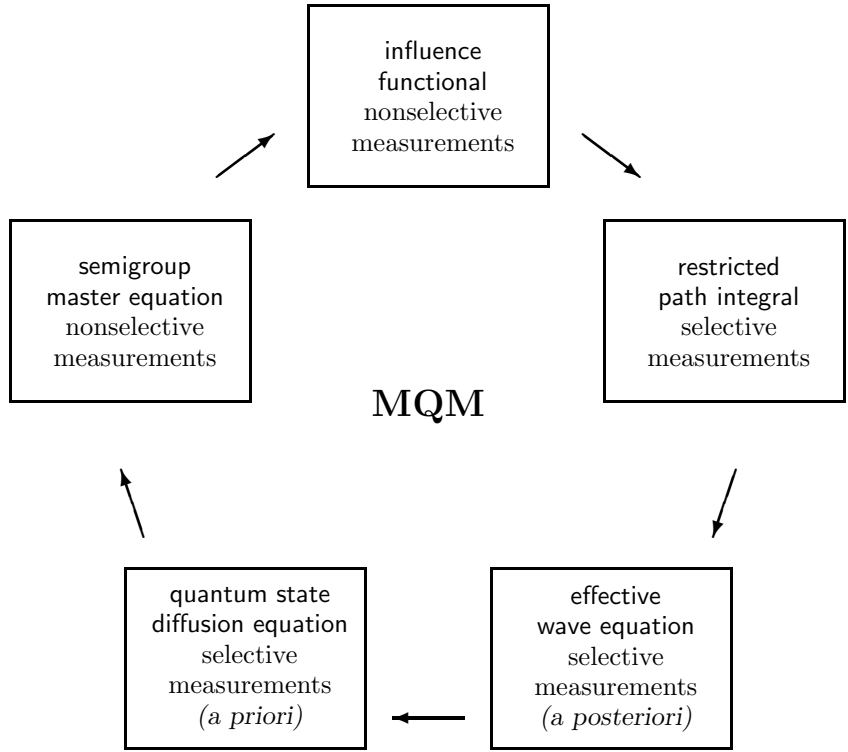
\begin{figure}
\begin{center}
\begin{picture}(314,314)(-40,-260)
\thicklines
\unitlength=0.75pt
\put(97.0,-45.0){\framebox(120,90){
\parbox{120pt}{\centering\normalsize 
{\sf influence \\ functional} \\ nonselective \\ measurements 
}}}
\put(-53.3,-153.5){\framebox(120,90){
\parbox{120pt}{\centering\normalsize 
{\sf semigroup \\ master equation} \\ nonselective \\ measurements 
}}}
\put(4.7,-329.0){\framebox(120,90){
\parbox{120pt}{\centering\normalsize 
{\sf quantum state \\ diffusion equation }\\  selective \\ measurements
\\ {\sl (a priori)}
}}}
\put(189.3,-329.0){\framebox(120,90){
\parbox{120pt}{\centering\normalsize 
{\sf effective \\ wave equation} \\ selective \\ measurements 
\\ {\sl (a posteriori)}
}}}
\put(246.3,-153.5){\framebox(120,90){
\parbox{120pt}{\centering\normalsize 
{\sf restricted \\ path integral} \\  selective \\ measurements 
}}}
\put(64.7,-30.0){\vector(4,3){14.4}}
\put(64.7,-30.0){\line(-4,-3){14.4}}
\put(249.3,-30.0){\line(-4,3){14.4}}
\put(249.3,-30.0){\vector(4,-3){14.4}}
\put(7.7,-205.5){\vector(-1,3){5.7}}
\put(7.7,-205.5){\line(1,-3){5.7}}
\put(306.3,-205.5){\line(1,3){5.7}}
\put(306.3,-205.5){\vector(-1,-3){5.7}}
\put(157.0,-314.0){\line(1,0){18.0}}
\put(157.0,-314.0){\vector(-1,0){18.0}}
\put(157.0,-157.0){\makebox(0,0){\Large \bf MQM}}
\end{picture}
\end{center}
\caption{Measurement Quantum Mechanics: connections among  
different approaches.}
\label{FIG1}
\end{figure}

The dynamics of a system interacting with an environment is conveniently
described in terms of a reduced density matrix operator $\hat \rho(t)$
obtained by tracing out the environment variables from the density matrix 
operator of the whole system + environment.
The unitary evolution of $\hat \rho(t)$ for the isolated system is 
modified to an irreversible one by the interaction with the environment. 
In the limit of a Markovian environment a dynamical law described by a
completely positive semigroup has been postulated \cite{GORINI,LINDBLAD}
resulting in the following
master equation for the reduced density matrix operator
\begin{equation}
{d \over dt} \hat \rho(t) = 
- {i \over \hbar} \left[ \hat H(t), \hat \rho(t) \right] 
+ {1\over2} \sum_\nu \left( 
\left[ \hat L_\nu(t) \hat \rho(t) , \hat L_\nu(t)^{\dag} \right] +
\left[ \hat L_\nu(t) , \hat \rho(t) \hat L_\nu(t)^{\dag} \right]
\right) 
\label{LINDBEQ}
\end{equation}
where $\hat H(t) = \hat H(\hat p, \hat q,t)$ is the Hamiltonian operator
for a general nonautonomous system and $\hat L_\nu(t)$ are the Lindblad
operators representing the influence of the environment on the system.

The above equation is thought to describe the general case of an
open quantum system. 
The evolution of a quantum system subjected to a measurement process 
is a particular case where the environment is the measurement apparatus 
and the Lindblad operators are proportional to the measured quantities.
If, for simplicity, we consider the measurement of a single observable 
represented by the Hermitian operator $\hat A(t) = \hat A(\hat p, \hat q,t)$, 
the corresponding Lindblad operator can be chosen as
$\hat L(t) = \hat L(t)^{\dag} = \kappa(t)^{1/2} \hat A(t)$. 
Note that we allow for an explicit time dependence of $\hat A(t)$ in order 
to include the case of general observables such as the continuous
quantum nondemolition ones \cite{CAVES}.
The function $\kappa(t)$ has dimensions $[\kappa]=[t^{-1} A^{-2}]$ 
and represents, as we shall see in the next section, the coupling 
of the measured system to the measurement apparatus.            
The measurement process is continuous in time and the measurement 
coupling is time dependent 
as requested in the description of a general experimental situation.

Due to the presence of commutators in the right hand side 
of (\ref{LINDBEQ}) the trace of the reduced density matrix operator  
is a conserved quantity which we assume to be unity
\begin{equation}
{\rm Tr} ~ \hat \rho(t) = 1 .
\label{TRACE}
\end{equation}
We shall see later, when introducing selective measurements, 
that the reduced density matrix operator $\hat\rho(t)$ corresponds 
to an incoherent mixture of pure states associated to selective processes.
In antithesis, the process described by $\hat\rho(t)$ is called nonselective
and Eq. (\ref{TRACE}) can be interpreted as a normalization relation for 
the probability distribution over the selective processes.
According to this interpretation, the result of a nonselective measurement, 
$\overline{a(t)}$, and its associated variance, $\Delta a(t)^2$,  
can be evaluated by the trace rules
\begin{equation}
\overline{a(t)} = {\rm Tr} \left[ \hat{A}(t) \hat\rho(t) \right] 
\label{AVENS}
\end{equation}
and 
\begin{equation}
\Delta a(t)^2 =
{\rm Tr} \left[ \left( \hat A(t) -\overline{a(t)} \right)^2 \hat \rho(t) 
\right],
\label{VAR}
\end{equation}
respectively. 
Here overlining is used to denote the statistical average over 
the selective measurement results $a(t)$ to be defined in the following 
together with the corresponding probability distribution.

The equivalence of the master equation (\ref{LINDBEQ}) to a density
matrix propagator expressed in terms of an influence functional
is our first step in the loop of Fig. 1.
By moving to the coordinate representation of the reduced density matrix 
operator 
\begin{equation}
\rho(q_1,q_2,t) =  \langle q_1 |~\hat \rho(t)~|q_2 \rangle 
\end{equation}
the corresponding evolution equation becomes the partial differential equation
\begin{eqnarray}
{\partial \over \partial t} \rho(q_1,q_2,t) = 
\Bigg[ &-&         
{i \over \hbar} H\left( -i \hbar {\partial \over \partial q_1},q_1,t \right) 
+{i \over \hbar} H\left( -i \hbar {\partial \over \partial q_2},q_2,t \right)   
\nonumber \\ 
&-& {1 \over 2} \kappa(t) 
\left[ A\left( -i \hbar {\partial \over \partial q_1},q_1,t \right) - 
	A\left( -i \hbar {\partial \over \partial q_2},q_2,t \right) \right]^2
\Bigg] \rho(q_1,q_2,t)  
\end{eqnarray}
which can be transformed into the integral equation
\begin{equation}
\rho(q_1'',q_2'',t'') = \int  dq_1' dq_2' ~
G(q_1'',q_2'',t'';q_1',q_2',t') ~ \rho(q_1',q_2',t') .
\label{INTEQ}
\end{equation}
The two-point Green function (density matrix propagator) $G$ has 
a phase-space path-integral representation which can be derived by
standard methods \cite{MENSKY94,CALDLEGG}.
For a small time interval $\Delta t$ one has 
\begin{equation}
\langle q_1''| \hat \rho(t'') | q_2'' \rangle = 
\langle q_1''| \hat \rho(t'' - \Delta t) | q_2'' \rangle +
\langle q_1''| {d \over dt} \hat \rho(t'' -\Delta t) 
|q_2'' \rangle \Delta t + {\cal O}(\Delta t^2).
\end{equation}
By using Eq. (\ref{LINDBEQ}) and inserting at the appropriate places 
the four identities 
\begin{equation}
\int dp_1^{(1)} |p_1^{(1)} \rangle \langle p_1^{(1)}|~
\int dq_1^{(1)} |q_1^{(1)} \rangle \langle q_1^{(1)}|~
\int dp_2^{(1)} |p_2^{(1)} \rangle \langle p_2^{(1)}|~ 
\int dq_2^{(1)} |q_2^{(1)} \rangle \langle q_2^{(1)}| 
\end{equation}
where 
\begin{equation}
\langle q | p \rangle = {1 \over \sqrt{2 \pi \hbar}} 
\exp \left( {i \over \hbar} p q \right)
\end{equation}
we get
\begin{eqnarray}
&&\rho(q_1'',q_2'',t'') = \int {dp_1^{(1)} \over 2 \pi \hbar} dq_1^{(1)}
\int {dp_2^{(1)} \over 2 \pi \hbar} dq_2^{(1)}
\exp \left( 
{i\over \hbar} \left[ p_1^{(1)} {q_1'' - q_1^{(1)} \over \Delta t}
-H \left(p_1^{(1)},q_1^{(1)},t'' - \Delta t \right) \right] \Delta t
\right. \nonumber \\
&&~-{1\over2} \kappa(t'' - \Delta t) 
\left[ A \left(p_1^{(1)},q_1^{(1)},t'' - \Delta t \right) -
A \left(p_2^{(1)},q_2^{(1)},t'' - \Delta t \right) \right]^2 \Delta t
\nonumber \\
&&~\left. - {i\over \hbar} \left[ p_2^{(1)} {q_2'' - q_2^{(1)} \over \Delta t}
-H \left(p_2^{(1)},q_2^{(1)},t'' - \Delta t \right) \right] \Delta t
\right)
\rho(q_1^{(1)},q_2^{(1)},t'' - \Delta t)  + {\cal O}(\Delta t^2). 
\end{eqnarray}
By iterating this relation $N$ times with $\Delta t = (t'' - t')/N$ 
and then taking the limit $N \to \infty$,
a functional measure arises
\begin{equation}
d[p]~ d[q]_{q',t'}^{q'',t''} =  \lim_{N \to \infty}
\prod_{n=1}^{N} {dp^{(n)} \over 2 \pi \hbar} 
\prod_{n=1}^{N-1}dq^{(n)}
\end{equation}
with boundary conditions imposed only to the $q(t)$ paths. 
By comparison with Eq. (\ref{INTEQ}) we conclude
\begin{eqnarray}
&&G(q_1'',q_2'',t'';q_1',q_2',t') =
\nonumber \\
&&~
\int d[p_1]~ d[q_1]_{q_1',t'}^{q_1'',t''} 
\int d[p_2]~ d[q_2]_{q_2',t'}^{q_2'',t''}~ 
\exp \left( {i\over \hbar} S[p_1,q_1] - {i\over \hbar} S[p_2,q_2]  
~- Z [p_1,q_1,p_2,q_2] \right)
\label{GREEN2}
\end{eqnarray}
where
\begin{equation}
S[p,q] = \int_{t'}^{t''} dt~ [p \dot q - H(p,q,t) ] 
\label{S}
\end{equation}
\begin{equation}
Z[p_1,q_1,p_2,q_2] = 
{1 \over 2} \int_{t'}^{t''} dt~ \kappa(t)
\left[ A(p_1,q_1,t) - A(p_2,q_2,t) \right]^2 .
\label{IPHASE}
\end{equation}
The effect of the measurement in the two-point Green function is 
represented by the functional $\exp(-Z)$ which reduces to the identity for 
$\kappa(t)=0$, i.e., in absence of a measurement process.
The functional $\exp(-Z)$ is the Feynman-Vernon influence functional  
\cite{FV} evaluated by tracing out the degrees of freedom of the environment
in the phase-space path-integral formulation of quantum mechanics.
As we shall see in the next section, the influence functional approach 
allows us to give an enlightening interpretation of the nonselective
measurement processes with an explicit expression of the parameter 
$\kappa(t)$. 

The second step in the equivalence loop of Fig. 1 is accomplished by
a formal manipulation of the influence functional \cite{MENSKY94}. 
By using the identity
\begin{eqnarray}
&&\exp \left( 
- {1 \over 2} \int_{t'}^{t''} dt~ \kappa(t)[A(p_1,q_1,t) - A(p_2,q_2,t)]^2 
\right) = \nonumber \\
&&~ \int d[a]~ \exp \left( 
- \int_{t'}^{t''} dt~ \kappa(t) [A(p_1,q_1,t) - a(t)]^2 
- \int_{t'}^{t''} dt~ \kappa(t) [A(q_2,p_2,t) - a(t)]^2 
\right) 
\label{IDENTITY}
\end{eqnarray}
where the functional measure arises by slicing the interval $[t',t'']$
into $N \to \infty$ subintervals at times $t^{(n)} = t'' - n \Delta t$
with $\Delta t = (t'' - t')/N$, i.e.,
\begin{equation}
d[a] =  \lim_{N \to \infty}
\prod_{n=1}^{N} da^{(n)} \sqrt{2 \kappa(t^{(n)}) \Delta t \over \pi} ,
\label{D[A]}
\end{equation}
the two-point Green function $G$ can be decomposed into a couple of 
one-point Green functions $G_{[a]}$
\begin{equation}
G(q_1'',q_2'',t'';q_1',q_2',t') = 
\int d[a]~ 
~G_{[a]}(q_1'',t'';q_1',t')  
~G_{[a]}(q_2'',t'';q_2',t')^*  
\label{G2ING1A}
\end{equation}
where
\begin{equation}
G_{[a]}(q'',t'';q',t') =
\int d[p] d[q]_{q',t'}^{q'',t''}
\exp \left(
{i \over \hbar} S[p,q] 
- \int_{t'}^{t''} dt~ \kappa(t) [A(p,q,t) - a(t)]^2 
\right) .
\label{G[A]}
\end{equation}
This one-point Green function is the phase-space generalization of the 
restricted path-integral approach originally proposed on physical grounds 
as a damping of the Feynman paths incompatible with
the measurement result \cite{ME1}.
Here the measurement result is the function $a(t)$ which, 
according to the interpretation of the functional measure $d[a]$,
is, in general, continuous but not differentiable.
If we assume that the measured system is in a pure state 
$|\psi(t') \rangle$ at the beginning of the measurement process, i.e., 
\begin{equation}
\rho(q_1',q_2',t') =
\langle q_1' | \psi(t') \rangle \langle \psi(t') | q_2' \rangle = 
\psi(q_1',t') \psi(q_2',t')^* ,
\end{equation}
Eq. (\ref{G2ING1A}) allows us to introduce a pure state 
$|\psi_{[a]}(t) \rangle$ which is the evolution at time $t>t'$ 
of the initial one $|\psi(t') \rangle$.
Indeed, the reduced density matrix can be thought as the functional integral
\begin{equation}
\rho(q_1,q_2,t) = \int d[a] ~\psi_{[a]}(q_1,t) \psi_{[a]}(q_2,t)^*  
\label{DECOA}
\end{equation}
with the wave function of the pure state defined by the one-point Green 
function $G_{[a]}$
\begin{equation}
\psi_{[a]}(q'',t'') = \int dq'~ G_{[a]}(q'',t'';q',t') ~ \psi(q',t'). 
\label{PSIAG}
\end{equation}
The decomposition (\ref{G2ING1A}) introduces the announced statistical 
description of a nonselective measurement process in terms of selective 
processes corresponding to different measurement results $a(t)$ and 
characterized by a certain probability distribution.
Due to Eq. (\ref{DECOA}) the conservation of the trace of the reduced 
density matrix operator gives
\begin{equation}
1 = \int d[a]~ \parallel \psi_{[a]}(t) \parallel^2 .
\label{PROBA}
\end{equation}
Therefore, $\parallel \psi_{[a]}(t) \parallel^2$ is the probability 
distribution for the selective process with measurement result $[a]$.
Once the probability distribution is known,  
the nonselective measurement result (\ref{AVENS})
and variance (\ref{VAR}) which, according to (\ref{DECOA}), 
are written also as 
\begin{equation}
\overline{a(t)} =
\int d[a] ~ \langle \psi_{[a]}(t) | \hat A(t) |\psi_{[a]}(t) \rangle 
\label{AVEA1}
\end{equation}
and
\begin{equation}
\Delta a(t)^2 =
\int d[a] ~ \langle \psi_{[a]}(t) | 
\left( \hat A(t) -  \overline{a(t)} \right)^2 |\psi_{[a]}(t) \rangle,
\label{VARA1}
\end{equation}
can be expressed in terms of the selective results $a(t)$ by
\begin{equation}
\overline{a(t)} =
\int d[a]~ \parallel \psi_{[a]}(t) \parallel^2 a(t) 
\label{AVEA2}
\end{equation}
and
\begin{equation}
\Delta a(t)^2 =  \lim_{\tau \to 0^{+}}
\int d[a]~ \parallel \psi_{[a]}(t) \parallel^2 
\left( a(t) - \overline{a(t)} \right)
\left( a(t-\tau) - \overline{a(t)} \right),
\label{VAR2}
\end{equation}
respectively.
The equivalence between (\ref{AVEA1}) and (\ref{AVEA2}) as well as
that one between (\ref{VARA1}) and (\ref{VAR2}) can be demonstrated   
explicitly by using the definition of the functional measure $d[a]$ 
and the functional expression of the wave function $\psi_{[a]}(q,t)$.
The prescription $\tau \to 0^+$ in (\ref{VAR2}) is used to avoid the
divergence which occurs by integrating the term $a(t)^2$.

The third step of our equivalence loop, namely the existence of a 
differential equation for the wave function $\psi_{[a]}(q,t)$, 
follows from the same standard method used for the reduced density matrix.
The explicit representation of the functional integral giving the one-point
Green function $G_{[a]}$ allows us to write the difference of the
wave function between two close times $t$ and $t + \Delta t$ up to 
terms ${\cal O}(\Delta t^2)$ and find the differential equation
\begin{equation}
{\partial \over \partial t} \psi_{[a]}(q,t) = 
\left[ -{i \over \hbar} 
H \left( -i \hbar {\partial \over \partial q},q,t \right)  
- \kappa(t) \left[ 
A \left( -i \hbar {\partial \over \partial q},q,t \right) - a(t)
\right]^2 
\right] \psi_{[a]}(q,t) .
\label{PSIA}
\end{equation}

The mathematical properties of this differential equation 
crucially depend on two possible ways of analyzing a selective
measurement process. 
A first possibility is to consider an analysis {\sl a posteriori}. 
The result of the measurement $a(t)$ is already known and one wants to
complete the quantum mechanical information on the measured system 
by evaluating the associated wave function $\psi_{[a]}(q,t)$.
In this case Eq. (\ref{PSIA}) is an effective wave equation
linear in the wave function. 
This equation is deterministic but not regular, in general, depending upon 
the nature of the function $a(t)$.
The effect of the measurement appears as an anti-Hermitian term
added to the Hamiltonian of the unmeasured system.
Due to this term the norm of the wave function $\psi_{[a]}(q,t)$
is not conserved in agreement with the probabilistic normalization 
(\ref{PROBA}).

A second possibility is to consider an analysis {\sl a priori}.
The result of the measurement $a(t)$ is to be predicted not 
deterministically but randomly with probability distribution 
$\parallel \psi_{[a]}(t) \parallel^2$.
Equation (\ref{PSIA}) becomes a stochastic differential equation 
nonlinear in $\psi_{[a]}(q,t)$.
This can be seen more explicitly by introducing an appropriate noise 
$\eta_{[a]}(t)$ which relates the selective result $a(t)$ to the
expectation value of the operator $\hat A(t)$ in the state
$| \psi_{[a]}(t) \rangle$.
Comparison of (\ref{AVEA1}) and (\ref{AVEA2}) allows us to write 
\begin{equation}
a(t) = 
{ \langle \psi_{[a]}(t) | \hat A(t) |\psi_{[a]}(t) \rangle  
\over \langle \psi_{[a]}(t) | \psi_{[a]}(t) \rangle }
+ \eta_{[a]}(t) 
\label{AINTOA}
\end{equation}
where $\eta_{[a]}(t)$ is defined by 
\begin{equation}
\int d[a]~ \parallel \psi_{[a]}(t) \parallel^2 \eta_{[a]}(t) = 0 .
\end{equation}
The nonlinearity of the stochastic equation for $\psi_{[a]}(q,t)$ is 
made clear by inserting (\ref{AINTOA}) into (\ref{PSIA}). 
This stochastic equation is also unconventional in the sense that 
it contains the square of the noise term $\eta_{[a]}(t)$.

It is evident that Eq. (\ref{PSIA}) is fully appropriate
in the {\sl a posteriori} analysis of a selective measurement process.
However, in the {\sl a priori} analysis one has  
to deal with an unconventional stochastic equation containing a complicated 
noise function defined in terms of a time-dependent probability distribution. 
It would be more desirable to have a standard stochastic equation with 
a white noise. 
This can be achieved, together with the fourth step in our equivalence loop, 
by a change of variable in the measurement result.
Analogously to Eq. (\ref{AINTOA}) we write
\begin{equation}
a(t) = {a}_{[\xi]}(t) + {\xi(t) \over 2 \sqrt{\kappa(t)}}
\label{AINTOAXI}
\end{equation}
where ${a}_{[\xi]}(t)$ is considered known for the moment 
and $\xi(t)$ is a noise whose characterization is obtained through the 
following considerations.
The two-point Green function already decomposed in (\ref{G2ING1A}) 
into a couple of
one-point Green functions $G_{[a]}$ with functional measure 
$d[a]$  can alternatively be decomposed into another couple of 
one-point Green functions $G_{[\xi]}$
\begin{equation}
G(q_1'',q_2'',t'';q_1',q_2',t')  
= \int d[\xi] 
~G_{[\xi]}(q_1'',t'';q_1',t')  
~G_{[\xi]}(q_2'',t'';q_2',t')^*  
\label{G2ING1XI}
\end{equation}
with Gaussian measure
\begin{equation}
d[\xi] =  \lim_{N \to \infty}
\prod_{n=1}^{N} d\xi^{(n)} \sqrt{\Delta t \over 2 \pi} 
\exp \left( -{1\over2} {\xi^{(n)}}^2  \Delta t \right)
\label{D[XI]}
\end{equation}
which makes $G_{[\xi]}$ exponentially linear in $\xi$
\begin{eqnarray}
&&G_{[\xi]}(q'',t'';q',t') =
\exp \left(\int_{t'}^{t''} dt~ {1\over4} \xi(t)^2 \right)
G_{[a]}(q'',t'';q',t') = 
\int d[p] d[q]_{q',t'}^{q'',t''}
\nonumber \\
&&~ \times \exp \left(
{i \over \hbar} S[p,q] 
- \int_{t'}^{t''} dt~ \kappa(t) 
\left[ [A(p,q,t)- {a}_{[\xi]}(t)]^2 - 
2 [A(p,q,t)- {a}_{[\xi]}(t)] {\xi(t) \over 2 \sqrt{\kappa(t)}} \right]
\right) .
\end{eqnarray}
With respect to the Gaussian measure $d[\xi]$, the noise 
$\xi(t)$ has average 
\begin{equation}
\overline{\xi(t)} = \int d[\xi]~\xi(t) = 0 
\label{XIMEDIO}
\end{equation}
and covariance
\begin{equation}
\overline{\xi(t_1)\xi(t_2)} = 
\int d[\xi]~ \xi(t_1)~\xi(t_2) = \delta(t_1-t_2) ,
\end{equation}
i.e., it is a white noise.
As in the previous case, the reduced density matrix can be written as a 
functional integral  
\begin{equation}
\rho(q_1,q_2,t) = \int d[\xi] ~\psi_{[\xi]}(q_1,t) \psi_{[\xi]}(q_2,t)^*  
\label{DECOXI}
\end{equation}
over pure states with wave functions $\psi_{[\xi]}(q,t)$ defined by 
\begin{equation}
\psi_{[\xi]}(q'',t'') = \int dq'~G_{[\xi]}(q'',t'';q',t') ~ \psi(q',t') .
\end{equation}
According to the expression of the Green function $G_{[\xi]}$,
the integral statement for $\psi_{[\xi]}(q,t)$ can be transformed
into an operatorial expression by the usual slicing procedure
\begin{eqnarray}
|\psi_{[\xi]}(t'') \rangle = 
\hat T \exp \Bigg( \int_{t'}^{t''} dt~ 
\bigg[
&-& {i \over \hbar} \hat H(\hat p,\hat q,t) 
- \kappa(t) \left[\hat A(\hat p,\hat q,t) - {a}_{[\xi]}(t) \right]^2  
\nonumber \\
&+& \sqrt{\kappa(t)} \left[ \hat A(\hat p,\hat q,t) - {a}_{[\xi]}(t) \right] 
\xi(t) \bigg] \Bigg) ~ |\psi_{[\xi]}(t') \rangle
\label{STOCSOL}
\end{eqnarray}
where $\hat T$ means chronological ordering.
By introducing the Wiener process $dw(t)=\xi(t) dt$ with Ito algebra
$dw(t)^2=dt$ and $dw(t)^{2+n}=0$ ($n > 0$),
Eq. (\ref{STOCSOL}) is recognized as the formal solution of the following 
Ito stochastic differential equation \cite{ARNOLD}
\begin{eqnarray}
d | \psi_{[\xi]}(t) \rangle &=& 
\left[ - {i \over \hbar} \hat H(\hat p,\hat q,t) 
- {1 \over 2} \kappa(t) \left[ \hat A(\hat p,\hat q,t)- {a}_{[\xi]}(t) 
\right]^2 \right]~ |\psi_{[\xi]}(t) \rangle ~dt 
\nonumber \\
&&~ 
+\sqrt{\kappa(t)} \left[ \hat A(\hat p,\hat q,t) - {a}_{[\xi]}(t) \right] 
~|\psi_{[\xi]}(t) \rangle~dw(t) .
\label{STOCEQ}
\end{eqnarray}
In (\ref{STOCEQ}) we still have to specify the function ${a}_{[\xi]}(t)$.
According to the decomposition (\ref{DECOXI}), the conservation  
of the trace of the reduced density matrix operator gives
\begin{equation}
1 = \int d[\xi] ~\langle \psi_{[\xi]}(t) | \psi_{[\xi]}(t) \rangle .
\end{equation}
Since the Gaussian measure $d[\xi]$ is normalized, 
up to a zero-average fluctuation we have 
$\langle \psi_{[\xi]}(t) | \psi_{[\xi]}(t) \rangle =1$.
The normalization of the state $|\psi_{[\xi]}(t) \rangle$ 
uniquely determines the expression of ${a}_{[\xi]}(t)$.
Indeed, by evaluating the Ito differential 
\begin{equation}
d ~ \langle \psi_{[\xi]}(t) | \psi_{[\xi]}(t) \rangle  =
2 \sqrt{\kappa(t)} ~ \langle \psi_{[\xi]}(t)
| \hat A(t) - {a}_{[\xi]}(t) | \psi_{[\xi]}(t) \rangle~
dw(t) + {\cal O}(dt^{3/2})
\end{equation}
we see that the norm of the state 
$|\psi_{[\xi]}(t) \rangle$ is conserved only if we impose
\begin{equation}
{a}_{[\xi]}(t) = 
\langle \psi_{[\xi]}(t) | \hat A(t) |\psi_{[\xi]}(t) \rangle . 
\label{NLXI}
\end{equation}
In this case the stochastic differential equation (\ref{STOCEQ})
becomes the nonlinear quantum state diffusion equation proposed
in \cite{DIOSI1,BELAVKIN1,GISIN2}.

Unlike the effective wave equation (\ref{PSIA}), the quantum state 
diffusion equation (\ref{STOCEQ}) is suitable for 
the {\sl a priori} analysis of the selective processes but 
awkward when dealing with the {\sl a posteriori} analysis.
Indeed, in the latter case one should guess the noise 
realization $\xi(t)$ giving rise, 
according to (\ref{AINTOAXI}) and (\ref{NLXI}), 
to the known value of the measurement result $a(t)$.
As in the decomposition of the reduced density matrix in terms of the
pure states $|\psi_{[a]}(t) \rangle$,
the nonselective measurement result (\ref{AVENS}) and variance 
(\ref{VAR}) can be expressed in terms of the pure states 
$|\psi_{[\xi]}(t) \rangle$ by 
\begin{equation}
\overline{a(t)} = 
\int d[\xi] ~ \langle \psi_{[\xi]}(t) | \hat A(t) |\psi_{[\xi]}(t) \rangle 
\label{AVEXI}
\end{equation}
and
\begin{equation}
\Delta a(t)^2 =
\int d[\xi] ~ \langle \psi_{[\xi]}(t) | 
\left( \hat A(t) - \overline{a(t)} \right)^2 |\psi_{[\xi]}(t) \rangle,
\label{VARXI}
\end{equation}
respectively.

As fifth and final step, the equivalence loop of Fig. 1 is closed by
regaining the starting differential equation for the reduced density matrix
operator
\begin{equation}
\hat \rho(t) = \overline{
| \psi_{[\xi]}(t) \rangle \langle \psi_{[\xi]} (t) | } 
\end{equation}
where overlining means functional integration with measure $d[\xi]$.
By using Ito algebra we get
\begin{eqnarray}
d \hat \rho(t) &=& 
\overline{  
d | \psi_{[\xi]}(t) \rangle ~ \langle \psi_{[\xi]}(t) | +
| \psi_{[\xi]}(t) \rangle ~ d \langle \psi_{[\xi]}(t) | + 
d | \psi_{[\xi]}(t) \rangle ~ d \langle \psi_{[\xi]}(t) | }
\nonumber \\
&=& \left(
- {i \over \hbar} \left[ \hat H(t), \hat \rho(t) \right] 
- {1 \over 2} \kappa(t)
\left[ \hat A(t), \left[ \hat A(t), \hat \rho(t) \right] \right]
\right) dt  + {\cal O}(dt^{3/2})
\end{eqnarray}
which is Eq. (\ref{LINDBEQ}) with the previously discussed choice for
the Lindblad operators.

\section{Influence functional from an informational environment}

Deeper insight into the problem of measurement in quantum mechanics
can be gained by completing the formal discussion of the previous 
section with general but explicit models for a measurement process.
For this purpose the influence functional approach is a quite 
appropriate one and some attempts have already been done in this direction 
\cite{NAMIOT,CAMIL,CAVES86,DOWKER,GELLMANN}.
Here we discuss the measurement process in terms of an informational 
environment made of particles linearly interacting with the measured 
system and in contact with a heat reservoir at fixed temperature. 
Each particle selectively measures the evolution of the system while  
the collection of them represents an apparatus performing
a nonselective measurement.
This model has two nice features.
Firstly, it is simple from a mathematical point of view and 
the corresponding influence functional can be evaluated exactly
\cite{CALDLEGG}.
Secondly, very clear and controllable approximations can be made on the
influence functional in order to reduce it to the formal 
expression given in the previous section.
These approximations are physically related to the fact that 
the informational environment can be defined as a many-body system in
which each particle,
still interacting quantum mechanically with the measured system, 
behaves classically from the point of view of the observer.
Only in this case the decision of the observer to look or 
not to look at the pointer of the instrument does not influence the result
of the measurement itself \cite{CINI}.

We start by considering the Hamiltonian of a measured system
interacting with an environment 
\begin{equation}
H_{tot} = H + H_{env} 
\label{HTOT}
\end{equation}
where $H=H(p,q,t)$ is the Hamiltonian of the measured system with
phase-space coordinates $p$ and $q$ 
and $H_{env}$ is the Hamiltonian of the interacting environment. 
The environment is made by different sets of particles with mass $M$
and phase-space coordinates $Q_{\nu n}$ and $P_{\nu n}$,
where $\nu$ labels the sets and $n$ the particles in each set.
The particles of each set $\nu$ interact linearly with an observable 
$A_{\nu}(p,q,t)$ of the system through a function $\lambda_{\nu}(t)$ 
which transduces a displacement of the observables $A_{\nu}$ 
into displacements of the coordinates $Q_{\nu n}$ 
\begin{equation} 
H_{env} = 
\sum_{\nu n} \left( 
\frac{ P_{\nu n}^2 }{ 2 M} +
\frac{M \omega_{\nu n}^2}{2} \left[ Q_{\nu n} - 
\lambda_{\nu}(t) A_{\nu}(p,q,t) \right]^2 \right).
\label{HENV}
\end{equation}
The interaction of the $n$th particle of the environment
with the $\nu$th observable is characterized by a proper angular
frequency $\omega_{\nu n}$.
In terms of spectral response we can say that each particle 
measures the Fourier component of $\lambda_{\nu}(t) A_{\nu}(p,q,t)$ 
at its proper angular frequency $\omega_{\nu n}$.

We assume that at time $t'$ the system is described by the density
matrix $\rho(q_1',q_2',t')$ and each particle of the environment is at 
thermal equilibrium with temperature $T$. 
Since the equilibrium value for the coordinates $Q_{\nu n}$ depends
on the coordinates $p$ and $q$, the assumption of thermal equilibrium 
at time $t'$ implies a correlation between the environment and the system
at the same instant of time \cite{PATRIARCA}.
The density matrix for the total system at time $t'$ can be written as 
\begin{equation}
\rho_{tot}(q_1',Q_1',q_2',Q_2',t') = 
\rho( q_1',q_2',t')~ 
\rho_{env} (\delta Q_1',\delta Q_2',t')
\end{equation}
where 
\begin{eqnarray}
&&\rho_{env} (\delta Q_1',\delta Q_2',t') =
\prod_{\nu n} 
\sqrt{ \frac{M \omega_{\nu n}}{\pi \hbar }
\tanh \left( {\hbar \omega_{\nu n} \over 2 k_B T} \right) } 
\nonumber \\
&&~ \times \exp \Bigg( 
- \frac{M \omega_{\nu n}} {2 \hbar }
\Bigg( 
{ \delta Q_{1\nu n}'^2 + \delta Q_{2\nu n}'^2 
\over \tanh(\hbar \omega_{\nu n}/k_BT) }
+ { 2 \delta Q_{1\nu n}' \delta Q_{2\nu n}' 
\over \sinh(\hbar \omega_{\nu n}/k_BT) }
\Bigg) \Bigg)
\label{RHOENV}
\end{eqnarray}
and
\begin{equation}
\delta Q_{\nu n}' = Q_{\nu n}' - \frac12 \lambda_\nu(t')
\left[ A_{\nu}(p_1',q_1',t') + A_{\nu}(p_2',q_2',t') \right] .
\end{equation}
The initial displacements $\delta Q_{\nu n}'$ are uniquely expressed 
in terms of the values assumed by the measured observable at the initial
points $p_1',q_1',t'$ and $p_2',q_2',t'$
by requiring translational and time reversal invariance \cite{PATRIARCA}. 

The steps required to get the expression of the influence functional are 
standard \cite{FV,CALDLEGG}.
At time $t'' > t'$ the reduced density matrix of the system,
obtained by tracing out the coordinates of the environment
in the total density matrix
\begin{equation}
\rho(q_1'' q_2'',t'') = 
\int dQ'' ~ \rho_{tot}(q_1'',Q'',q_2'',Q'',t'') ,
\end{equation}
can be obtained by propagating the initial density matrix 
$\rho(q_1',q_2',t')$ 
\begin{equation}
\rho(q_1'' q_2'', t'') = 
\int dq_1' dq_2'  
~G( q_1'',q_2'',t'';q_1',q_2',t')~
\rho(q_1',q_2',t')  
\end{equation}
by the two-point Green function
\begin{eqnarray}
&&G(q_1'',q_2'',t'';q_1',q_2',t') = 
\nonumber \\ &&
   \int d[p_1] d[q_1]_{q_1',t'}^{q_1'',t''} 
   \int d[p_2] d[q_2]_{q_2',t'}^{q_2'',t''}
\exp \left( \frac{i}{\hbar} S[p_1,q_1] - \frac{i}{\hbar} S[p_2,q_2] \right)
F[p_1,q_1,p_2,q_2]
\end{eqnarray}
where $S[p,q]$ is given by (\ref{S}) and the influence functional $F$ is
\begin{eqnarray}
&& F[p_1,q_1,p_2,q_2] =
\int dQ'' \int dQ_1' dQ_2'
\int d[P_1] d[Q_1]_{Q_1',t'}^{Q'',t''}
\int d[P_2] d[Q_2]_{Q_2',t'}^{Q'',t''}
\nonumber \\&&~\times
\exp \left( 
  \frac{i}{\hbar} S_{env}[P_1,Q_1,A_1] 
- \frac{i}{\hbar} S_{env}[P_2,Q_2,A_2] \right)
\rho_{env} (\delta Q_1',\delta Q_2',t')
\label{INF}
\end{eqnarray}
with 
\begin{equation}
S_{env}[P,Q,A] = \int_{t'}^{t''} dt~ \left(
\sum_{\nu n} P_{\nu n} \dot Q_{\nu n} - H_{env}(P,Q-\lambda A,t) \right) . 
\end{equation}
In the above formulas the notation $P$, $Q$, $\delta Q$ is a shortening for 
$\{P_{\nu n}\}$, $\{Q_{\nu n}\}$, $\{\delta Q_{\nu n}\}$ and $\lambda$, $A$ 
for $\{\lambda_\nu\}$, $\{A_\nu\}$.
The integrations in (\ref{INF}) are Gaussian and can be performed giving
\begin{eqnarray}
&&F[p_1,q_1,p_2,q_2] = \prod_{\nu} \exp \Bigg( 
- \int_{t'}^{t''} dt \int_{t'}^{t} ds ~
\lambda_\nu(t) \left[ A_{\nu}(p_1,q_1,t) - A_{\nu}(p_2,q_2,t) \right]
\Big( \alpha_\nu(t-s) \lambda_\nu(s) 
\nonumber \\ &&\times 
\left[ A_{\nu}(p_1,q_1,s) - A_{\nu}(p_2,q_2,s) \right]  
+ i \beta_\nu(t-s) {d \over ds} \left( \lambda_\nu(s)
\left[ A_{\nu}(p_1,q_1,s) + A_{\nu}(p_2,q_2,s) \right] \right)
\Big) \Bigg) 
\label{GFI}
\end{eqnarray}
where the two kernels
\begin{equation}
\alpha_\nu(t-s) = \frac{M}{2 \hbar} \sum_n \omega_{\nu n}^3
\coth \left( \frac{\hbar \omega_{\nu n}}{2 k_B T} \right) 
\cos[\omega_{\nu n} (t-s)]
\label{FLU1}
\end{equation}
and 
\begin{equation}
\beta_\nu(t-s) = \frac{M}{2 \hbar} \sum_n \omega_{\nu n}^2 
\cos[\omega_{\nu n} (t-s)]
\label{DIS}
\end{equation}
describe fluctuation and dissipation phenomena, respectively \cite{CALDLEGG}.

Now we turn the attention to the requirement that the environment is 
informational, i.e., classical with respect to the observer so that 
the readout of information through the coordinates $Q$ has an objective value.
Since a classical system is one whose quantized structure can not be 
appreciated we must impose the thermal fluctuations to be large in comparison 
to the quanta of the environment
\begin{equation}
k_B T \gg \hbar \omega_{\nu n} .
\label{CLASSIC}
\end{equation}
In this case we have
\begin{equation}
\alpha_{\nu}(t,s) =
\frac{M k_B T}{\hbar^2} 
\sum_n \omega_{\nu n}^2 
\cos[\omega_{\nu n} (t-s)] .
\label{FLU2}
\end{equation}

Memory effects in the fluctuation and dissipation kernels are 
an inessential complication which can be avoided by 
assuming an ensemble of particles in the environment with
a continuous spectrum of frequencies. 
If, for simplicity, we choose the same frequency density 
$dN_\nu /d\omega = \Omega / \pi \omega^2$ for each set $\nu$
we get
\begin{equation}
\sum_n \omega_{\nu n}^2 \cos[\omega_{\nu n} (t-s)] \simeq
\int_{0}^{\infty} d\omega {d N_\nu \over d \omega}  
\omega^2 \cos[\omega (t-s)] =
\Omega \delta (t-s)  .
\end{equation}

Of course, the above assumption of continuous frequency spectrum 
implies that the condition (\ref{CLASSIC}) can not be satisfied in 
the whole frequency range $[0,+\infty[$ by a finite temperature.
Therefore, we should assume the condition (\ref{CLASSIC}) to be valid 
only in a finite range $[0,\omega_{max}]$ which contains the most 
significant part of the Fourier spectrum of $\lambda_\nu(t) A_\nu(p,q,t)$ 
and is relevant in monitoring the measured observables.  
In this case the condition (\ref{CLASSIC})
allows us to neglect the dissipation term, proportional to (\ref{DIS}), 
with respect to the fluctuation one, proportional to (\ref{FLU2}). 
In conclusion, the influence functional for an informational environment  
can be written as
\begin{equation}
F[p_1,q_1,p_2,q_2] = \prod_{\nu} \exp \left(
- \frac12 \int_{t'}^{t''} dt ~\kappa_\nu(t)
\left[ A_{\nu}(p_1,q_1,t) - A_{\nu}(p_2,q_2,t) \right]^2 
\right)
\label{FVERNON}
\end{equation}
with 
\begin{equation}
\kappa_{\nu}(t) = \lambda_{\nu}(t)^2
\frac{2 M \Omega k_B T} {\hbar^2} .
\label{KAPPONE}
\end{equation}
Equations (\ref{FVERNON}) and (\ref{KAPPONE}) generalize to more 
than one monitored quantity the expression 
of the functional already introduced in (\ref{IPHASE}) for
nonselective measurements.
Note that $\kappa_{\nu}(t)$ can be written as 
$\kappa_\nu(t)=\lambda_\nu(t)^2 / \sigma^2 \tau$
where $\sigma^2 = \hbar / 2 M \Omega$ and $\tau=\hbar / k_B T$. 
The two parameters $\sigma$ and $\tau$ can be made indefinitely small 
by increasing the density of meters (particles) in the informational 
environment proportional to $\Omega$, or their temperature $T$, respectively.
These characteristic length and time play the role of similar parameters
introduced {\sl ad hoc} in the spontaneous localization \cite{GHIRIWE}
and in the restricted path-integral \cite{MENSKYBOOK} approaches.
  
We close this section by evaluating the measurement result arising
as a readout from the informational environment and its
corresponding variance. 
The observer reads the result of the nonselective measurement of the
$\nu$th observable by looking at a pointer which responds to
the coordinates of the measurement apparatus.
This response is a weighted sum over the proper frequencies of the 
informational environment and is characterized by a normalized response 
function $\zeta_{\nu n}$ such that the pointer displacement and 
its variance are
\begin{equation}
P_\nu(t) = \sum_n \zeta_{\nu n} \overline{Q_{\nu n}(t)} 
\end{equation}
\begin{equation}
\Delta P_\nu(t)^2 = \sum_n \zeta_{\nu n} \Delta Q_{\nu n}(t)^2  
\end{equation}
where 
\begin{equation}
\overline{Q_{\nu n}(t)} = 
{\rm Tr} \left[ \hat Q_{\nu n} \hat\rho_{tot}(t) \right]
\label{TR1}
\end{equation}
\begin{equation}
\Delta Q_{\nu n}(t)^2 = 
{\rm Tr} \left[
\left(\hat Q_{\nu n} - \overline{Q_{\nu n}(t)} \right)^2
\hat\rho_{tot}(t) \right] .
\label{TR2}
\end{equation}

Due to the condition (\ref{CLASSIC}) the thermal relaxation time of the
measurement apparatus turns out to be much smaller than the characteristic 
timescales of the measured system.
This allows us to use an adiabatic approximation for the total density 
matrix operator and factorize it as
\begin{equation}
\hat\rho_{tot}(t) \simeq \hat\rho(t) ~ \hat\rho_{env}(t)
\end{equation}
where $\hat\rho(t)$ is the reduced density matrix operator 
of the system which takes into account the influence of the informational
environment and $\hat\rho_{env}(t)$ is the density matrix operator 
of the informational environment at thermal equilibrium around the 
instantaneous value of the measured observables.
In the representation of the environment coordinates $\hat\rho_{env}(t)$
has matrix elements
\begin{eqnarray}
&& \langle Q_1 | \hat\rho_{env} (t) |Q_2 \rangle=
\prod_{\nu n} 
\sqrt{ \frac{M \omega_{\nu n}}{\pi \hbar }
\tanh \left( {\hbar \omega_{\nu n} \over 2 k_B T} \right) } 
\nonumber \\
&&~ \times \exp \Bigg( 
- \frac{M \omega_{\nu n}} {2 \hbar }
\Bigg( 
{ \delta \hat Q_{1\nu n}(t)^2 + \delta \hat Q_{2\nu n}(t)^2 
\over \tanh(\hbar \omega_{\nu n}/k_BT) }
+ { 2 \delta \hat Q_{1\nu n}(t) \delta \hat Q_{2\nu n}(t) 
\over \sinh(\hbar \omega_{\nu n}/k_BT) }
\Bigg) \Bigg)
\end{eqnarray}
which are operators with respect to the system coordinates through
the displacements
\begin{equation}
\delta \hat Q_{\nu n}(t) = Q_{\nu n}(t) - \frac12 \lambda_\nu(t)
\left[ \hat A_{\nu}(\hat p_1,\hat q_1,t) + 
\hat A_{\nu}(\hat p_2,\hat q_2,t) \right] .
\end{equation}
In this case the traces in (\ref{TR1}) and (\ref{TR2}) contain Gaussian 
integrals over the environment coordinates which can be performed giving
\begin{equation}
{\rm Tr} \left[ \hat Q_{\nu n} \hat \rho_{tot}(t) \right] =
\lambda_\nu(t) {\rm Tr} \left[ \hat A_\nu(t) \hat \rho(t) \right]
\end{equation}
\begin{equation}
{\rm Tr} \left[ \hat Q_{\nu n}^2 \hat \rho_{tot}(t) \right] =
{\hbar \over 2 M \omega_{\nu n}} 
\coth \left( {\hbar \omega_{\nu n} \over 2 k_B T} \right)
+ \lambda_\nu(t)^2 {\rm Tr} \left[ \hat A_\nu(t)^2 \hat \rho(t) \right] .
\end{equation}
In the limit (\ref{CLASSIC}) and for a continuous spectrum of frequencies
with the previously chosen density we have
\begin{equation}
P_\nu(t) = \lambda_\nu(t) \overline{ a_\nu(t)}
\label{P}
\end{equation}
\begin{equation}
\Delta P_\nu(t)^2 = 
\int d\omega ~ \zeta_\nu(\omega) {d N_\nu \over d \omega}
{k_B T \over M \omega^2}
+ \lambda_\nu(t)^2 \Delta a_\nu(t)^2 
\label{DP}
\end{equation}
where $\overline{ a_\nu(t)}$ and $\Delta a_\nu(t)^2$ are given by 
(\ref{AVENS})  and (\ref{VAR}), respectively.

Up to the transduction factor $\lambda_\nu(t)$ and $\lambda_\nu(t)^2$, 
respectively, Eq.s  (\ref{P}) and (\ref{DP}) are the result of the measured 
observable $A_\nu$ and its variance read from the pointer.
The result of the measurement is the nonselective outcome of the observable 
$A_\nu$ obtained from measurement quantum mechanics.
The measurement variance is the sum of 
a classical variance associated to the measurement apparatus 
and the quantum variance associated to the measured system. 
The nonselective result and variance (\ref{P}) and (\ref{DP})
can be expressed in terms of selective processes.
The decomposition of $\overline{ a_\nu(t)}$ and $\Delta a_\nu(t)^2$
has already been discussed in the previous section. 
Concerning the decomposition of the variance
associated to the measurement apparatus we note that the nonselective 
measurement performed by the informational environment is achieved by 
summing over the frequencies of the particles in the environment. 
Therefore, the contribution to the measurement variance given by the 
selective process at frequency $\omega$ is $k_BT/M\omega^2$ 
as expected from the equipartition theorem.

Equation (\ref{DP}) sets the definition of classical and quantum 
measurements.
The first ones are those characterized by a dominance of the 
variance associated to the measurement apparatus,
i.e., by a dominance of the thermal (or Brownian) noise.
In the second ones the quantum variance of the measured system
is larger than the thermal noise.
This second case promises a richer variety of experimental scenarios  
because the measurement variance depends on the preparation of the 
measured system and/or the strength of the coupling to the measurement 
apparatus.

\section{Experiments on quantum Zeno effect}

In many experimental situations one is dealing with an average 
of measurements on an individual quantum system each time prepared 
in the same initial state or a single measurement on an ensemble of 
independent identical quantum systems with the same initial conditions.
In both cases or in a combination of them averaged measurement results,
instead of individual measurement results, are actually registered as 
outcome of the experiment. 
According to the discussion of section II we have two ways for 
theoretically reproducing the outcome of such an experiment.
We can consider {\sl a priori} selective measurements and obtain 
the experiment outcome by averaging the corresponding selective
results in the sense of Eq. (\ref{AVEXI}).
We can also consider a nonselective measurement and evaluate the
experiment outcome directly from Eq. (\ref{AVENS}).
The choice between the two methods may depend on the particular problem
one is faced to.
The method based on a nonselective process is more direct but it
is based upon the solution of a master equation for the reduced density matrix
operator which could be much more difficult than solving many times the 
the diffusion state equation (\ref{STOCEQ}) and averaging the selective 
measurement results.
In this section we will deal with an experimental situation which can be 
described by a simple 2 by 2 density matrix.
The choice of the direct method based on nonselective measurements 
is, therefore, the natural one.

Let us consider a system with time-independent Hamiltonian and 
discrete energy spectrum
\begin{equation}
\hat H | n \rangle = E_n | n \rangle .
\end{equation}
The evolution of the system subjected both 
to an external time-dependent perturbation $\hat V(t)$ and 
to a continuous nonselective measurement of the observable represented 
by the operator $\hat A$ is given by
\begin{equation}
{d \hat \rho(t) \over dt} =
- {i \over \hbar} \left[ \hat H + \hat V(t), \hat \rho(t) \right] 
- {1 \over 2} \kappa(t)
\left[ \hat A, \left[ \hat A, \hat \rho(t) \right] \right] .
\end{equation}
An interesting situation is attained in the case the external perturbation
$\hat V$ stimulates the system to make transitions among the unperturbed 
levels $n$ and the occupancy of some of these levels is measured.
At what extent does the measurement disturb the stimulated transitions?
Eventually, inhibition of the stimulated transitions due to the 
occupancy measurement occurs and one has an example of what is 
called quantum Zeno effect \cite{KHA,MIS1,MIS2,COOK,KRAU,SUDB,BLANCHARD}.

From a theoretical point of view a two-level system is the simplest one 
for studying the interplay between the effects of the measurement process 
and of the external perturbation.
This situation has been also experimentally investigated \cite{ITA}.
An ensemble of about 5000 $^9Be^+$ ions was stored in 
a Penning trap.
Two hyperfine levels of the ground state of $^9Be^+$, created by
a static magnetic field and hereafter called levels 1 and 2, were
driven by a radiofrequency, resonant between levels 1 and 2, turned on
for $T=256$ ms.
The amplitude of the radiofrequency was adjusted to make   
the initially vanishing occupancy of level 2  to be unity at time $T$ 
in absence of other disturbances.
During the radiofrequency pulse, $n$ optical pulses of length $\tau=2.4$ ms 
and frequency equal to the transition frequency between level 1 and a 
third level 3 were also applied.
The number of photons emitted in the spontaneous transition 
$3 \to 1$, the transition $3 \to 2$ being forbidden, 
was roughly proportional to the occupancy of level 1. 
The optical pulses acted, therefore, as a measurement of the occupancy of 
level 1.
The occupancy at time $T$ of level 1 was observed to be frozen near its 
initial unity value, i.e., the stimulated transition $1 \to 2$ in the period 
$T$ to be inhibited, proportionally to the number $n$ of optical pulses. 

Before trying a direct interpretation of the experiment \cite{ITA},
let us consider the case of a continuous measurement of the
occupancy of level 1 in a two-level system subjected to 
stimulated transitions.   
In the representation of the unperturbed Hamiltonian $\hat H$
where $\rho_{nm} = \langle n | \hat \rho | m \rangle$ 
we assume a perturbation $\hat V$ with matrix elements
$V_{11}=V_{22}=0$ and $V_{12}=V_{21}^*=V_0 e^{i\omega (t-t_0)}$ with
$V_0$ real and $\omega=(E_2-E_1)/{\hbar} + \delta \omega$.
In the same representation the matrix elements of the measured occupancy 
of level 1 are $A_{11}=1$ and $A_{12}=A_{21}=A_{22}=0$.
The master equation for the reduced density matrix operator then gives
\begin{eqnarray}
\dot{\rho}_{11}(t)&=&
- {i\over \hbar} \left[ V_{12} \rho_{21}(t) - \rho_{12}(t) V_{21} \right]
\nonumber \\
\dot{\rho}_{22}(t)&=&
- {i\over \hbar} \left[ V_{21} \rho_{12}(t) - \rho_{21}(t) V_{12} \right]
\\ 
\dot{\rho}_{12}(t)&=&
\left[ -{i\over\hbar} (E_1-E_2) - {\kappa \over 2} \right] \rho_{12}(t)
- {i\over \hbar} \left[ V_{12} \rho_{22}(t) - \rho_{11}(t) V_{12} \right]
\nonumber
\end{eqnarray}
and $\rho_{21}(t)=\rho_{12}(t)^*$. 
The measurement coupling $\kappa$ is assumed constant.
By summing and subtracting the first two equations we find
\begin{eqnarray}
\dot{\rho}_{11}(t)+\dot{\rho}_{22}(t)&=&0 \nonumber\\
\dot{\rho}_{22}(t)-\dot{\rho}_{11}(t)&=&\frac{4}{\hbar} 
{\rm Im} \left( \rho_{12}(t) V_{21} \right) \\
\dot{\rho}_{12}(t)&=&
\left[ -{i\over\hbar} (E_1-E_2) - {\kappa \over 2} \right] \rho_{12}(t)
- {i\over \hbar} V_{12} \left( \rho_{22}(t) - \rho_{11}(t) \right) .
\nonumber
\end{eqnarray}
The first equation is the conservation of 
the trace of the reduced density matrix operator in the case of
a nonselective measurement, $\rho_{11}(t)+\rho_{22}(t)=1$.
By defining
\begin{equation}
\rho_{12}(t) = \exp \left( i \omega (t-t_0) \right) 
\left[ \alpha(t) + i \beta(t) \right] 
\qquad
\rho_{22}(t) - \rho_{11}(t) = \gamma(t) 
\end{equation}
with $\alpha$, $\beta$ and $\gamma$ real, the other two equations give
\begin{eqnarray}
\dot{\alpha}(t) &=& - {\kappa \over 2} \alpha(t) + \delta \omega \beta(t) 
\nonumber\\
\dot{\beta}(t) &=& - {\kappa \over 2} \beta(t) - {\omega_R \over 2} \gamma(t) 
- \delta \omega \alpha(t) \\
\dot{\gamma}(t)&=&2 \omega_R \beta(t) \nonumber
\end{eqnarray} 
where we have introduced the Rabi angular frequency 
\begin{equation}
\omega_R = 2 {V_0 \over \hbar}  .
\label{OMEGAR}
\end{equation}
The above system has a simple solution for $\delta\omega=0$ (resonance) 
and in this case we get
\begin{eqnarray}
\rho_{11}(t)&=&
\frac12 - \frac12 e^{-{1\over4} \kappa t }
\Bigg[ (\rho_{22}(0)-\rho_{11}(0)) 
\left(\cos(wt)+\frac{\kappa}{4w}\sin(wt)\right) 
\nonumber \\ &&+
{\rm Im} \left( \rho_{12}(0) e^{i \omega t_0} \right)
\frac{2\omega_R}{w} \sin(wt) \Bigg]
\label{RHO11} \\
\rho_{12}(t)&=& 
e^{ -\frac{i}{\hbar}(E_1-E_2)t -{1\over4} \kappa t -i\omega t_0 } 
\Bigg[ {\rm Re} \left(\rho_{12}(0)  e^{i \omega t_0}\right)
e^{ -{1\over4} \kappa t } 
\nonumber \\ &&+
i {\rm Im} \left( \rho_{12}(0) e^{i \omega t_0}\right) 
\left( \cos(wt) - \frac{\kappa}{4w} \sin(wt) \right)
- i \left( \rho_{22}(0)-\rho_{11}(0)\right)
\frac{\omega_R}{2w}\sin(wt) \Bigg] 
\label{RHO12}
\end{eqnarray}
where
\begin{equation}
w = \sqrt{ \omega_R^2 -{1\over16} \kappa^2 } .
\end{equation}
The angular frequency $w$ coincides with the Rabi angular frequency $\omega_R$ 
when $\kappa=0$.
In this case the effect of the measurement disappears and the system 
oscillates between levels 1 and 2 with angular frequency $\omega_R$. 
In the opposite limit of strong measurement coupling the frequency $w$ 
is imaginary and an overdamped regime is achieved in which transitions 
are inhibited. 
The border between the two regimes is at $w=0$ corresponding 
to a critical measurement coupling
\begin{equation}
\kappa_{crit}=4 \omega_R = 8 { V_0 \over \hbar }.
\label{KCRIT}
\end{equation}

The behavior of the measured occupancy of level 1 which, according
to Eq. (\ref{AVENS}), turns out to be $\rho_{11}(t)$
is shown in Fig. 2 for different values of the measurement coupling
starting from initial conditions $\rho_{11}(0)=1$ and $\rho_{12}(0)=0$.
The transition between the Rabi-like regime and the Zeno-like regime 
is marked by the disappearing of the oscillatory behavior at 
$\kappa=\kappa_{crit}$.

The quantum variance associated to the nonselective measurement 
of level 1 can be evaluated according to Eq. (\ref{VAR}) and is
$\rho_{11}(t) \rho_{22}(t)$.
It vanishes in the limit of strong measurement coupling and 
oscillates with angular frequency $\omega_R$ between 0 and $1/4$ 
in the opposite limit $\kappa \to  0$.
At the critical measurement coupling $\kappa=\kappa_{crit}$,
after a short transient of the order of $\omega_R^{-1}$, 
the quantum variance approaches the constant and maximum 
value $(1/2)^2$. 

It is worth to note that an identical behavior of the reduced density
matrix operator is obtained if we consider $\hat H$ as the measured quantity. 
In this case Eq.s (\ref{RHO11}) and (\ref{RHO12}) still
hold with the substitution 
$\kappa \to \kappa_E (E_2-E_1)^2$ where $\kappa_E$ is the measurement 
coupling associated to the measurement of $\hat H$.
This observation allows us to compare Fig. 2 with Fig. 1 of 
Ref. \cite{OPT} where the quantum Zeno effect for the same 
two-level system investigated here was analyzed {\sl a posteriori}
in the case of a single selective measurement of energy.
As expected, quantitative differences between nonselective measurements
and single selective measurements are obtained but in both cases Zeno 
inhibition occurs for strong measurement coupling.
The behavior shown in Fig. 2 agrees also with that found in
\cite{GAMIL,GAWIMIL} where the quantum Zeno effect is analyzed 
within the quantum trajectory approach \cite{GRIFF}. 

\begin{figure}
\centerline{\hbox{\psfig{figure=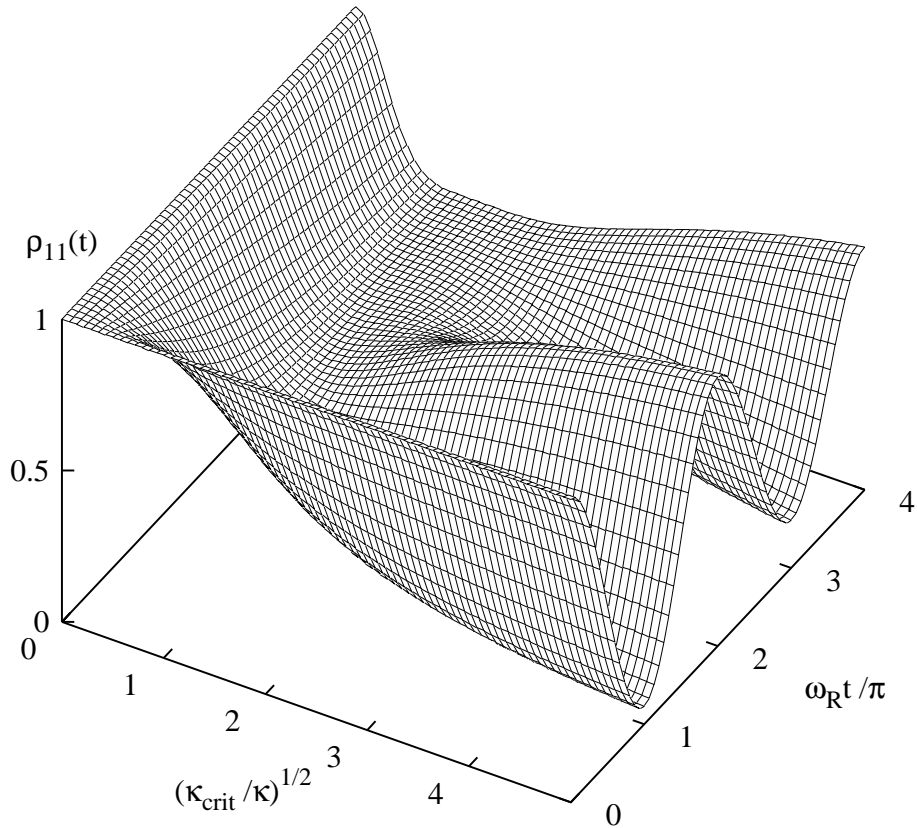,width=15cm,angle=270}}}
\caption{Measurement result $\rho_{11}(t)$ during the nonselective
measurement of occupancy of level 1 in a two-level system
simultaneously driven by a resonant perturbation
as a function of the adimensional quantity $\kappa_{crit}/\kappa$.
The system is prepared in level 1 at time $t=0$.}
\label{FIG2}
\end{figure}
The discussion relative to Fig. 2 can be made general.
In the limit $\kappa \to 0$ the influence of the measurement disappears
and ordinary quantum mechanics is recovered.
In the limit $\kappa \to \infty$ the influence of the measurement dominates
and the average measurement result (\ref{AVENS}) as well as the corresponding
variance (\ref{VAR}) are frozen into the values assumed at the beginning 
of the measurement.
In the case of Fig. 2 these values are $\rho_{11}(0)=1$ and 
$\rho_{11}(0) \rho_{22}(0)=0$, respectively.
Moreover, in the limit $\kappa \to \infty$ the off-diagonal elements of
the reduced density matrix operator vanish and a classical behavior is 
obtained.

In order to recover the results of the experiment \cite{ITA} the previous
analysis for a continuous nonselective measurement with $\kappa$ constant 
must be generalized to a series of measurement pulses spaced by intervals
of no measurement. 
According to the experimental procedure we consider $n$ nonselective
measurements of the occupancy of level 1 with coupling $\kappa$ 
during the intervals $[jT/n-\tau,jT/n]$, $j=1, \ldots,n$, 
with $\tau=2.4$ ms and $T=256$ ms.
During the remaining part of the interval $[0,T]$ the system is
subjected only to the radiofrequency perturbation at 
$\omega= 2 \pi \times 320.7$ MHz whose amplitude $V_0$ is fixed 
by the condition $\omega_R=\pi/T$ that gives, in absence of measurement, 
$\rho_{22}(T)=1$ if $\rho_{22}(0)=0$.
Since the radiofrequency is resonant $\delta \omega=0$.
The time evolution of the reduced density matrix operator corresponding 
to the described process is obtained by successive iterations of 
(\ref{RHO11}) and (\ref{RHO12}) with the same formulas evaluated 
for $\kappa=0$.
The theoretical probability for the transition $1 \to 2$ at the end of the 
interval $T$
\begin{equation}
P_{1\to 2}^{th}(n,\kappa) = 1 - \rho_{11}(T)
\end{equation}
can be compared with the corresponding experimental data
$P_{1\to 2}^{exp}(n)$ available for $n=1, 2, 4, 8, 16, 32, 64$.
The uncertainty for the experimental transition probabilities
is estimated to be $\Delta P=0.02$ \cite{ITA}.

In Fig. 3 we show the sum of the squared differences between 
the theoretical and experimental transition probabilities normalized to 
the experimental uncertainties, the $\chi^2$, as a function the 
phenomenological parameter $\kappa / \kappa_{crit}$. 
In the same figure we show also the probability $Q(\chi^2|\nu)$ that 
the observed value for the chi-square should exceed the value 
$\chi^2(\kappa / \kappa_{crit})$ by chance.
In our case the number of degrees of freedom is $\nu=7-1$.
The two arrows indicate the chi-square value obtained by the same authors
of \cite{ITA} using a theoretical
model based on the instantaneous von Neumann collapse 
without (higher value) or with (lower value)
corrections for the finite duration of the 
measurement pulses, the effect of optical pumping from level 2 
to level 1 and the measured value of $\omega_R$.
Depending upon the statistical confidence level we adopt, 
from Fig. 3 we see that values of $\kappa/\kappa_{crit} \gtrsim 10^2$ 
are required to fit properly the experimental data.
The minimum value of $\chi^2$ obtained from measurement quantum mechanics
slightly differs from the $\chi^2$ obtained on the basis of the 
von Neumann postulate including experimental corrections.

\begin{figure}
\centerline{\hbox{\psfig{figure=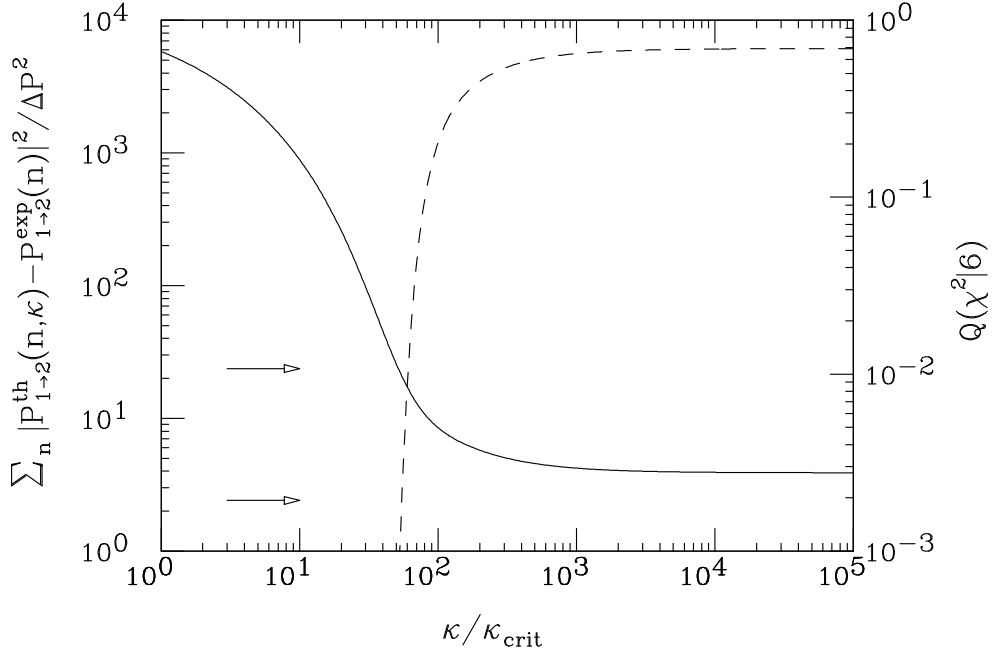,width=15cm,angle=90}}}
\caption{Behavior of 
$\chi^2 = \sum_n \left| P_{1\to 2}^{th}(n,\kappa)-
P_{1\to 2}^{exp}(n) \right|^2 / \Delta P^2$ as a function of 
$\kappa / \kappa_{crit}$ in fitting the experiment \protect\cite{ITA}.
The dashed line is the probability $Q(\chi^2|6)$ that the 
chi-square should exceed the value $\chi^2(\kappa / \kappa_{crit})$ by chance.
The two arrows indicate the chi-square value obtained in \protect\cite{ITA} 
by a model based on instantaneous von Neumann collapse 
without (higher value) or with (lower value) 
corrections for finite measurement-pulse duration,
optical pumping and measured value of $\omega_R$.}
\label{FIG3}
\end{figure}
It is neither surprising nor exciting
that measurement quantum mechanics is able to explain the experimental 
results of Ref. \cite{ITA} in terms of a strong Zeno inhibition.
Not surprising because measurement quantum mechanics contains more naive
approaches to the problem of quantum measurements as the von Neumann
postulate, for instance, which was already shown to reproduce 
the experimental results.
Not exciting because strong Zeno inhibition as well as full Rabi
oscillations are two trivial extreme regimes. 
However, measurement quantum mechanics tells us that another interesting 
and unexplored regime exists.
It is the regime which occurs when the measurement coupling is comparable 
to the critical value.
In this case a strong competition between stimulated 
transitions and measurement inhibition takes place as clearly
recognized from Fig. 2 and also from the analysis of
the corresponding quantum variance, $\rho_{11}\rho_{22}$,  
which gets maximum at $\kappa = \kappa_{crit}$.
Measurement quantum mechanics suggests us also how to 
explore this regime.
In order to make $\kappa / \kappa_{crit} \sim 1$ we should 
decrease $\kappa$ and/or increase $\kappa_{crit}$ with respect to the 
values used in \cite{ITA}.
According to the discussion of section III $\kappa$ can be decreased
by reducing the density of particles in the informational environment.
In the experiment \cite{ITA} the ``particles'' of the informational 
environment are the excited modes of the photon vacuum.    
Their density can be reduced by lowering the intensity of the optical
radiation acting as a measurement probe.
On the other hand, Eq. (\ref{KCRIT}) shows that $\kappa_{crit}$  
can be increased by means of the amplitude $V_0$ of the perturbation
which stimulates the Rabi oscillations.
In the experiment \cite{ITA} this amounts to increase the intensity 
of the radiofrequency radiation.  
Of course, by increasing $V_0$ the Rabi angular frequency increases too
so that the condition $\omega_R=\pi/T$ used in \cite{ITA} 
can be maintained only by decreasing the period $T$ and, consequently,
the duration of the measurement pulses. 
The possibility to vary the strength of the coupling is also related to the 
opportunity of looking for decoherence effects in the same experiment.
Indeed, from Eq. (\ref{RHO12}) we see that the off-diagonal elements
of the density matrix vanish exponentially with time constant 
$\tau_{dec}=4/\kappa$. 
For an appropriate measurement coupling, i.e., a proper intensity
of the optical radiation, $\tau_{dec}$ could be 
experimentally accessible by optical homodyne tomography techniques 
\cite{VOGEL,SMITHEY,DARIANO} and  decoherence phenomena could be investigated.

\section{Conclusions}

Measurement quantum mechanics has been introduced by showing the 
equivalence among formalisms developed with the aim of
including the effect of the measurement on the dynamics of a quantum system. 
The theory contains one parameter, the measurement coupling, 
for each measured observable.

The measurement couplings can be expressed in terms of the detailed 
properties of a measurement apparatus which extracts information 
from the measured system in objective way, i.e., 
independently of the presence of an observer looking at the pointer
of the apparatus.
The definition of quantum and classical measurements then 
emerges naturally according to the dominance, in the uncertainty 
of the pointer, of the classical fluctuations due to the apparatus or 
of the quantum fluctuations due to the measured system. 

Alternatively, one can introduce the measurement couplings as 
phenomenological parameters to be inferred from comparison 
with experimental data, as in the example of the quantum Zeno effect. 

Ordinary quantum mechanics is obtained 
as a limit case for vanishing measurement couplings.
In the opposite limit of infinite measurement couplings the 
dynamics of the measured system is frozen and classical.
Both limits are not interesting, the first one because there is no
measurement and the second one because there is no quantum dynamics to measure.
Interesting physics is in between and measurement quantum mechanics
is a tool for designing experiments in this intermediate regime.
	   
\acknowledgments 
We are grateful to 
G. Alli, M. Cini, G. Jona Lasinio, L. Morato, M. Patriarca and L. Viola 
for profitable discussions. 
This work was supported in part by INFN Iniziativa Specifica RM6.
%
%

%
%


\begin{references}

\bibitem{VONNEUM}{\sc J. von Neumann}, 
``Mathematical Foundations of Quantum Mechanics,''
Princeton University Press, Princeton, 1955.

\bibitem{WHEE}{\sc J. A. Wheeler and W. H. Zurek (Eds.)},
``Quantum Theory and Measurement,''  
Princeton University Press, Princeton, 1983.

\bibitem{BUSCH}{\sc P. Busch, P. J. Lahti, and P. Mittelstaedt}, 
``The Quantum Theory of Measurement,'' Springer-Verlag, Berlin, 1991.

\bibitem{GREEN}{\sc D. M. Greenberger (Ed.)},
``New Techniques and Ideas in Quantum Measurement Theory,'' 
New York Academy of Sciences, New York, 1986.

\bibitem{BRAGIN}{\sc V. B. Braginsky and F. Ya. Khalili},   
``Quantum Measurements,'' {\sc K. S. Thorne (Ed.)},  
Cambridge University Press, Cambridge, 1992.

\bibitem{GORINI}{\sc V. Gorini, A. Kossakowski, and E. C. G. Sudarshan}, 
{\sl J. Math. Phys.} {\bf 17} (1976), 821.

\bibitem{LINDBLAD}{\sc G. Lindblad}, 
{\sl Comm. Math. Phys.} {\bf 48} (1976), 119.

\bibitem{CARMICHAEL}{\sc H. J. Carmichael}, ``An Open System Approach
to Quantum Optics,'' Springer-Verlag, Berlin, 1993.

\bibitem{ZUREK}{\sc W. H. Zurek}, {\sl Phys. Rev. D} {\bf 24} (1981), 1516;
{\sl ibidem} {\bf 26} (1982), 1862.

\bibitem{GHIRIWE}{\sc  G. C. Ghirardi, A. Rimini, and T. Weber}, 
{\sl Phys. Rev. D} {\bf 34} (1986), 470.

\bibitem{DIOSI2}{\sc L. Di\'osi}, 
{\sl Phys. Rev. A} {\bf 40} (1989), 1165.

\bibitem{BALAMPRO}{\sc A. Barchielli, L. Lanz, and G. M. Prosperi}, 
{\sl Nuovo Cimento B} {\bf 72} (1982), 121.

\bibitem{LUDWIG}{\sc G. Ludwig}, ``An Axiomatic Basis for Quantum Mechanics,'' 
vol. 2, Springer-Verlag, Berlin, 1985.

\bibitem{DGHP} For a connection between localization and decoherence see
{\sc L. Di\'osi, N. Gisin, J. Halliwell, and I. C. Percival}, 
{\sl Phys. Rev. Lett.} {\bf 74} (1995), 203.

\bibitem{BARCHIELLI}{\sc A. Barchielli}, 
{\sl Nuovo Cimento B} {\bf 74} (1983), 113;
{\sl Phys. Rev. A} {\bf 34} (1986), 1642.

\bibitem{DIOSI1}{\sc L. Di\'osi}, {\sl Phys. Lett. A} {\bf 129} (1988), 419.

\bibitem{DIOSI1B}{\sc L. Di\'osi}, {\sl Phys. Lett. A} {\bf 132} (1988), 233.

\bibitem{LUDER}{\sc G. L\"uders}, 
{\sl Ann. Phys.} (Leipzig) {\bf 8} (1951), 322.

\bibitem{GISIN1}{\sc N. Gisin}, {\sl Phys. Rev. Lett.} {\bf 52} (1984), 1657;
{\sl ibidem} {\bf 53} (1984), 1776.

\bibitem{BELAVKIN1}{\sc V. P. Belavkin}, 
{\sl Phys. Lett. A} {\bf 140} (1989), 355;
V. P. Belavkin and P. Staszewski, {\sl Phys. Lett. A} {\bf 140} (1989), 359.

\bibitem{GISIN2}{\sc N. Gisin and I. C. Percival}, 
{\sl J. Phys. A} {\bf 25} (1992), 5677; 
{\sl Phys. Lett. A} {\bf 167} (1992), 315.

\bibitem{BELAVKIN2}{\sc V. P. Belavkin}, 
{\sl J. Phys. A} {\bf 22} (1989), L1109.

\bibitem{BARBE}{\sc A. Barchielli and V. P. Belavkin}, 
{\sl J. Phys. A} {\bf 24} (1991), 1495.

\bibitem{FEY}{\sc R. P. Feynman}, {\sl Rev. Mod. Phys.} {\bf 20} (1948), 367.

\bibitem{ME1}{\sc M. B. Mensky}, {\sl Phys. Rev. D} {\bf 20} (1979), 384;
{\sl Zh. Eksp. Teor. Fiz.} {\bf 77} (1979), 1326
[{\sl Sov. Phys. JETP} {\bf 50} (1979), 667].

\bibitem{GOMES}{\sc G. A. Golubtsova and M. B. Mensky}, 
{\sl Int. J. Mod. Phys. A} {\bf 4} (1988), 2733.

\bibitem{OPT}{\sc R. Onofrio, C. Presilla, and U. Tambini}, 
{\sl Phys. Lett. A} {\bf 183} (1993), 135. 

\bibitem{MENSKYBOOK}{\sc M. B. Mensky}, 
``Continuous Quantum Measurements and Path Integrals,'' chapter 4,  
Institute of Physics Publishing, Bristol, Philadelphia, 1993.

\bibitem{NAMIOT}{\sc A. Konetchnyi, M. B. Mensky, and V. Namiot},
{\sl Phys. Lett. A} {\bf 177} (1993), 283.

\bibitem{TPO}{\sc U. Tambini, C. Presilla, and R. Onofrio}, 
{\sl Phys. Rev. A} {\bf 51} (1995), 967.

\bibitem{MEOP1}{\sc M. B. Mensky, R. Onofrio, and C. Presilla}, 
{\sl Phys. Lett. A} {\bf 161} (1991), 236.

\bibitem{MEOP2}{\sc M. B. Mensky, R. Onofrio, and C. Presilla}, 
{\sl Phys. Rev. Lett.} {\bf 70} (1993), 2825.

\bibitem{CALO}{\sc T. Calarco and R. Onofrio}, 
{\sl Phys. Lett. A} {\bf 198} (1995), 279.

\bibitem{DIOSI4}{\sc L. Di\'osi}, ``Selective continuous quantum 
measurements: Restricted path integrals and wave equations,'' 
electronic archives ref.: quant-ph/9501009.                   

\bibitem{MENSKY94}{\sc M. B. Mensky}, {\sl Phys. Lett. A} {\bf 196} (1994), 159.

\bibitem{FV}{\sc R. P. Feynman and F. L. Vernon}, 
{\sl Ann. Phys.} {\bf 24} (1963), 118.

\bibitem{CAMIL}{\sc C. M. Caves and G. J. Milburn}, 
{\sl Phys. Rev. A} {\bf 36} (1987), 5543.

\bibitem{CAVES86}{\sc C. M. Caves}, {\sl Phys. Rev. D} {\bf 33} (1986), 1643;
{\sl ibidem} {\bf 35} (1987), 1851.

\bibitem{CINI}{\sc M. Cini}, {\sl Nuovo Cimento B} {\bf 73} (1983), 27.

\bibitem{CALDLEGG}{\sc A. O. Caldeira and A. J. Leggett}, 
{\sl Physica A} {\bf 121} (1983), 587.

\bibitem{ITA}{\sc W. M. Itano, D. J. Heinzen, J. J. Bollinger, 
and D. J. Wineland}, {\sl Phys. Rev. A} {\bf 41} (1990), 2295. 

\bibitem{CAVES}{\sc C. M. Caves, K. S. Thorne, R. P. Drever, V. D. Sandberg, 
and M. Zimmermann}, {\sl Rev. Mod. Phys.} {\bf 52} (1980), 341.

\bibitem{ARNOLD}{\sc L. Arnold}, ``Stochastic Differential Equations: 
Theory and Applications,'' John Wiley \& Sons, New York, 1974.

\bibitem{DOWKER}{\sc H. F. Dowker and J. J. Halliwell},
{\sl Phys. Rev. D} {\bf 46} (1992), 1580.

\bibitem{GELLMANN}{\sc M. Gell-Mann and J. B. Hartle},
{\sl Phys. Rev. D} {\bf 47} (1993), 3345.

\bibitem{PATRIARCA}{\sc M. Patriarca}, ``Statistical correlations in the 
oscillator model of quantum dissipative systems,'' 
submitted to {\sl Nuovo Cimento B}.

\bibitem{KHA}{\sc L. A. Khalfin}, 
{\sl Zh. Eksp. Teor. Fiz.} {\bf 33} (1957), 1371 
[{\sl Sov. Phys. JETP} {\bf 6} (1958), 1053]; 
{\sl Zh. Eksp. Teor. Fiz. Pis'ma Red.} {\bf 8} (1968), 106  
[{\sl JETP Lett.} {\bf 8} (1968), 65].

\bibitem{MIS1}{\sc B. Misra and E. C. G. Sudarshan}, 
{\sl J. Math. Phys.} {\bf 18} (1977), 756.

\bibitem{MIS2}{\sc C. B. Chiu, E. C. G. Sudarshan, and B. Misra}, 
{\sl Phys. Rev. D} {\bf 16} (1977), 520.

\bibitem{COOK}{\sc R. J. Cook}, {\sl Phys. Scr.} {\bf T 21} (1988), 49.

\bibitem{KRAU}{\sc K. Kraus}, {\sl Found. of Phys.} {\bf 11} (1981), 547.

\bibitem{SUDB}{\sc A. Sudbery}, {\sl Annals Phys.} {\bf 157} (1984), 512.

\bibitem{BLANCHARD}{\sc Ph. Blanchard and A. Jadczyk}, 
{\sl Phys. Lett. A} {\bf 183} (1993) 272.

\bibitem{GAMIL}{\sc M. J. Gagen and G. J. Milburn}, 
{\sl Phys. Rev. A} {\bf 47} (1993), 1467.

\bibitem{GAWIMIL}{\sc M. J. Gagen, H. M. Wiseman, and G. J. Milburn}, 
{\sl Phys. Rev. A} {\bf 48} (1993), 132.

\bibitem{GRIFF}{\sc R. B. Griffits}, {\sl J. Stat. Phys.} {\bf 36} (1984), 219;
{\sc P. Zoller, M. Marte, and D. F. Walls}, 
{\sl Phys. Rev. A} {\bf 35} (1987), 198;
{\sc H. J. Carmichael, S. Singh, R. Vyas, and P. R. Rice}, 
{\sl Phys. Rev. A} {\bf 39} (1989), 1200.

\bibitem{VOGEL}{\sc K. Vogel and H. Risken}, 
{\sl Phys. Rev. A} {\bf 40} (1989), 2847.

\bibitem{SMITHEY}{\sc D. T. Smithey, M. Beck, M. G. Raymer, and A. Faridani}, 
{\sl Phys. Rev. Lett.} {\bf 70} (1993), 1244.

\bibitem{DARIANO}{\sc G. M. D'Ariano, C. Machiavello, and M. G. A. Paris}, 
{\sl Phys. Rev. A} {\bf 50} (1994), 4298.


\end{references}
\end{document}